\documentclass[preprint,12pt]{aastex}

\usepackage{amsmath}
\usepackage{amssymb}
\usepackage{graphicx}

\begin{document}

\newcommand{\diaz}{N$_2$H$^+$}

\shorttitle{Hyperfine emission}
\shortauthors{Keto and Rybicki}

\title{Modeling molecular hyperfine line emission}

\author{Eric Keto and George Rybicki }
\affil{Harvard-Smithsonian Center for Astrophysics,}
\affil{60 Garden St., Cambridge, MA 02138}

\begin{abstract}
In this paper we discuss two approximate methods previously 
suggested for modeling hyperfine spectral line
emission for molecules whose collisional transitions rates
between hyperfine levels are unknown. Hyperfine structure is seen in
the rotational spectra of many commonly observed molecules such as
HCN, HNC, NH$_3$, \diaz , and C$^{17}$O. The intensities of these
spectral lines 
can be modeled by
numerical techniques such as $\Lambda-$iteration that 
alternately solve the equations of statistical equilibrium 
and the equation of radiative transfer.
However, these
calculations require knowledge of both the radiative and collisional
rates for all transitions.  For most commonly
observed radio frequency spectral lines, only the 
net collisional rates between rotational levels are known.
For such cases, two approximate methods have been suggested.
The first method, hyperfine statistical equilibrium ({\it HSE}), distributes
the hyperfine level populations 
according to their statistical weight,
but allows the population of the rotational states
to depart from local thermodynamic equilibrium (LTE).
The second method, the {\it proportional} method approximates
the collision rates between the hyperfine levels
as fractions
of the net rotational rate apportioned according to the statistical
degeneracy of the final hyperfine levels.
The second method is able to model non-LTE
hyperfine emission. 
We compare simulations of \diaz\ hyperfine lines made
with approximate and
more exact rates 
and find that satisfactory results are
obtained. 
\end{abstract}


\section{Introduction} \label{introduction}

The rotational spectra of many commonly observed molecules such as
HCN, HNC, NH$_3$, \diaz , and C$^{17}$O exhibit hyperfine structure
from the splitting of the rotational energy levels by electric quadrupole
and magnetic dipole interactions induced by the nuclear moments of
atoms such as N or $^{17}$O with non-zero spin. 
Hyperfine lines reduce the effective
optical depth of the rotational transition
by spreading the emission out over a wider bandwidth.
Because estimates of the
density, temperature, and molecular abundance
depend on the optical depth,
the hyperfine structure should be taken into account
in analyzing spectral line observations
Properly treated, the hyperfine structure is quite useful.
The observed relative intensities of pairs of hyperfine lines 
constrain the optical depth 
independently of the molecular abundance and independently of the
spatial coupling of the telescope beam with the cloud structure (beam
filling factor). In contrast, optical depth determination from
the brightness ratios of spectral lines of isotopologues such as $^{12}$CO and $^{13}$CO
requires knowledge of the isotopic abundance ratios, and furthermore the lines
may be at sufficiently different frequencies that the observing beam may
be differently coupled to the structure of the cloud.

Numerical techniques such as $\Lambda-$iteration that 
alternately solve the equations of statistical equilibrium to determine the
level populations and the equation of radiative transfer to determine
the mean radiation field are able to predict line intensities over a broad
range of conditions
including varying temperature and density
and non-LTE excitation.
However, these
calculations require knowledge of both the radiative and collisional
rates for all transitions.  This presents a problem in the case of the
hyperfine lines. For most molecules, the radiative rates, Einstein
$A_{ij}$, are known for all the transitions 
including hyperfine transitions, but the collisional
rates are known only as the net rates between rotational levels.
These net rates represent the weighted sum of the rates of 
all the individual hyperfine transitions between the rotational levels.
Collisional
rates between the individual hyperfine levels themselves
have been calculated for only three molecules: HCN (Monteiro \&
Stutzki 1986), NH$_3$ (Chen, Zhang \& Zhou 1998), and \diaz \ (Daniel
et al 2005), and even then for only a limited number of hyperfine
levels.

Two approximations have been suggested for modeling the emission from
molecules with unknown hyperfine collisional rates.
The first approximation, 
"hyperfine statistical equilibrium" ({\it HSE}), assumes that the
the hyperfine levels within each rotational level are 
populated in proportion to their statistical 
weights \citep{Keto1990, Keto2004}. The second approximation, 
the {\it proportional} approximation,
assumes that the collisional rates
between the individual
hyperfine levels are proportional to the total rate between their
rotational levels and the statistical degeneracy
of the final hyperfine level of the transition \citep{GB1991,Daniel2006}.

In this paper, we discuss and evaluate these two approximations,
and compute sample \diaz\ spectra from each method. Because the
collisional rates for the hyperfine transitions of \diaz\
are known \citep{Daniel2005} we can compare \diaz\ 
spectra produced using the approximate collisional rates of the
{\it proportional} method 
against spectra produced using the ``exact" rates determined from
the numerical quantum mechanical calculations. 
We also show how the collisional rates for the elastic ($\Delta J = 0$)
rotational transitions may be determined by extrapolation from
the inelastic rates. 
These elastic rates are required in the {\it proportional} approximation
in order to determine
the collisional rates for hyperfine transitions 
within the same rotational state.
However, the elastic rates are generally not included in compilations
of calculated rate coefficients.

\section{First Approximation: Hyperfine Statistical Equilibrium}

Within a non-LTE model of radiative transfer such as the
Approximate (or Accelerated) Lambda Iteration (ALI) or 
Monte Carlo methods, we can approximate 
the splitting of hyperfine emission in a simple way
even if we only know the net collisional rates 
for the rotational transitions. 
The approximation is based on the difference in the
magnitude of 
the energies  of the hyperfine and rotational transitions.
The hyperfine levels of
molecules that emit in the
millimeter radio spectrum are typically separated by
energies in the milli-Kelvin range whereas the
separation between rotational levels are several tens to
hundreds of Kelvin.  Therefore, 
the hyperfine levels may sometimes be populated approximately in
statistical equilibrium even if the rotational levels
are not.  
For example, observations of \diaz\ often show brightness
ratios between hyperfine lines that depart from
LTE at only 10\% of the line brightness \citep{Tafalla2004}.
In such cases, for some observational purposes,
the assumption of hyperfine statistical equilibrium ({\it HSE})
may be adequate.
If not, the {\it HSE} method is not appropriate.

There are several advantages of this {\it HSE} approximation.
It automatically
takes into account overlapping emission from the
hyperfine lines. Therefore it may
be implemented with a simple alteration of 
the standard ALI algorithm (Rybicki \& Hummer 1991)
rather than the more
complex ALI algorithm for overlapping lines (Rybicki \& Hummer 1992). 
Another advantage is that only the rotational lines
require radiative transfer modeling.
There are always fewer rotational lines than hyperfine lines.
Of course, a
rotational transition split by hyperfine structure 
requires a larger bandwidth, but 
even so, the computational time is
much faster than modeling
all the individual hyperfine lines.

We implement this method in ALI starting 
the same way as 
for molecules without hyperfine splitting. We solve
the equations of statistical equilibrium 
to determine the populations in the rotational
levels
using the radiative and collision rates between 
the rotational levels, and an estimate of the mean 
radiation field from the previous iteration (RH91 equation 2.27). 

\begin{eqnarray}\label{eq:SEHSE}
&&\sum_{J>J'} n_{J} A_{JJ'}(1-\bar{\Lambda}_{JJ'}) 
- (n_{J'}B_{J'J} - n_{J}B_{JJ'})\bar{J}^{eff}_{JJ'} 
\nonumber \\
&+&\sum_{J'>J} n_{J'} A_{J'J}(1-\bar{\Lambda}_{JJ'}) 
- (n_{J}B_{JJ'} - n_{J'}B_{J'J})\bar{J}^{eff}_{JJ'} 
\nonumber \\
&+&\sum_{J'}(n_{J}C_{JJ'} - n_{J'}C_{J'J}) = 0\hfill
\end{eqnarray}
where the effective mean radiation field, $\bar{J}^{eff}_{JJ'}$,
is defined in ALI as,
\begin{equation}\label{eq:JbareffHSE}
\bar{J}^{eff}_{JJ'} = \bar{J}_{JJ'} - \bar{\Lambda}_{JJ'}S_{JJ'}
\end{equation}

Here the initial and final rotational levels are denoted by subscripts
$J$ and $J'$, the Einstein A and B rate coefficients by $A$ and $B$, 
the collisional rate coefficients by $C_{JJ'}$, the mean intensity 
by $\bar{J}_{JJ'}$, and the level
populations by $n_J$.
The approximate or accelerated 
Lambda-iteration operator is
$\bar{\Lambda}_{JJ'}$
where the overbar indicates the average
over frequency. (In RH91 this operator is denoted 
$\bar{\Lambda}^\ast_{\ell\ell'}$.)
The source function, $S_{JJ'}$, is the usual source function
between rotational levels,
\begin{equation}\label{eq:sourceHSE}
S_{JJ'} = {{j_{JJ'}}\over{\alpha_{JJ'}}}
\end{equation}
where $j_{JJ'}$ and $\alpha_{JJ'}$ are the emissivity
and opacity defined below.

To determine the approximate hyperfine line emission
we assume that the population of
each rotational level is divided among its hyperfine
states 
according to their statistical weights,
\begin{equation}\label{eq:HSEpops}
     n_{JH}= {g_{JH} \over g_J} n_J,    
\end{equation}
where $g$ is the statistical
weight and $H$ denotes a hyperfine level.

We do not need to
actually compute or store the 
populations of the hyperfine levels.
The assumption of
hyperfine statistical equilibrium is equivalent to the assumption
that the spectral line profile function
of the rotational transition
including the hyperfine structure is the sum of
the spectra of the individual hyperfine lines with
the same relative intensities as in optically thin emission.
Because these relative intensities depend only on the
the dipole matrix elements of the hyperfine radiative
transitions, we compute the composite profile function
once and then
replace
the simple line profile function of the rotational transition
with the
composite profile function 
everywhere in the calculation.

For example, if the line profile function of an unsplit rotational 
transition would be described by a particular
function, $\phi_1(\nu)$, for example a Gaussian, the 
line profile function with hyperfine splitting
would be the
sum of copies of the same Gaussian, 
one for each of the individual hyperfine transitions from
$JH$ to $J'H'$, each weighted by the individual relative line
intensity, $R_{JJ'HH'}$ and shifted in frequency according
to the energy difference, $\nu_{JJ'HH'}$, of the hyperfine splitting,
\begin{equation}\label{eq:sumprofile}
\phi_{JJ'}(\nu) = \sum_{HH'} \phi_{JJ'HH'}(\nu)
\end{equation}
and
\begin{equation}\label{eq:phiH}
\phi_{JJ'HH'}(\nu) = \phi_1(\nu+\nu_{JJ'HH'})R_{JJ'HH'}.
\end{equation}
Here $JHJ'H'$ means $JH \rightarrow J'H'$.
If the relative intensities, $R_{JJ'HH'}$ are normalized, then so is the 
composite profile function,
\begin{equation}
\int \phi_{JJ'}(\nu) d\nu = 1.
\end{equation}

The relative intensities of the
hyperfines and their frequency shifts are very simply calculated for
molecules with
only one atom
with an interacting nuclear moment. This case requires
only
the angular momentum quantum numbers of the initial and final
states using
the same formulas for 
the relative intensities and frequencies of atomic fine structure lines 
(equations 6-6a,b in Townes \& Schawlow 1956). 
This follows from the analogy between a transition
that changes the angular momentum of a molecule without altering its
nuclear spin and a transition that changes the orbital angular
momentum of the electrons in an atom without changing the electron
spins. The hyperfine relative intensities and frequencies in
molecules with two mutually interacting atoms such as \diaz\  
may be determined with a perturbation technique (Townes \&
Schawlow 1956) but generally
numerical techniques (Pickett 1991) are required for high precision.

Using the composite
profile function,
the line emissivity and opacity including the hyperfine lines
can now be calculated from the level populations, $n_J$,
of the rotational states.
The line emissivity is,
\begin{equation}\label{eq:emissivityJ}
j_{JJ'} = {{h\nu}\over{4\pi}}n_JA_{JJ'}\phi_{JJ'}(\nu)
\end{equation}
and the line opacity is,
\begin{equation}\label{eq:opacityJ}
\alpha_{JJ'} = {{h\nu}\over{4\pi}}(n_{J'}B_{J'J} - n_JB_{JJ'})\phi_{JJ'}(\nu)
\end{equation}
The mean radiation field is also computed with
the composite line profile function,
\begin{equation}\label{eq:JbarJ}
\bar{J}_{JJ'} = \int_\nu I_{JJ'}(\nu) \phi_{JJ'}(\nu) d\nu
\end{equation}
Similarly, 
the ALI  
operator is, 
\begin{equation}\label{eq:aliJ}
\bar{\Lambda}_{JJ'} = \int_\Omega d\Omega \int_\nu d\nu \bigg(1-\exp{\big(-\tau_{JJ'C}(\nu)\big)} \bigg)
r_{JJ'C}
\phi_{JJ'}
\end{equation}
if we use just the diagonal term. 
Here the optical depth, $\tau_{JJ'C}$ 
is defined as the line opacity (equation \ref{eq:opacityJ}) 
plus the continuum
opacity times the pathlength,$L$,
\begin{equation}\label{eq:tau}
\tau_{JJ'C} = (\alpha_{JJ}(\nu) + \alpha_{C}) L.
\end{equation}
The factor, $r_{JJ'C}$ is defined as in RH91 eqn 2.91
\begin{equation}\label{eq:r291J}
r_{JJ'C} = {{\alpha_{JJ'}(\nu)}\over{\alpha_{JJ'}(\nu)}+\alpha_C}
\end{equation}
and $\alpha_C$ is the opacity of the continuum.
If there is no continuum, then $\alpha_C = 0$ and $r_{JJ'C} = 1$.

We can now calculate the radiation along a ray in the usual
way by dividing the ray into cells, $i$, with constant
excitation temperature and density,
\begin{equation}\label{eq:RT}
I_{JJ'}^{i+1}(\nu) = I_{JJ'}^i(\nu) \exp{(-\tau_{JJ'}(\nu))}
+ S_{JJ'C}(1-\exp{(-\tau_{JJ'}(\nu))}
\end{equation}
where
the source function including the continuum is defined,
\begin{equation}\label{eq:sourceC}
S_{JJ'C}(\nu) = {{1}\over{2k}} 
{{ j_{JJ'}(\nu) + j_{C} } \over { \alpha_{JJ'}(\nu) + \alpha_{C} }}
\end{equation}
Equation \ref{eq:RT} shows that the relative intensities of the
individual hyperfine lines in the spectrum $I_{JJ'}(\nu)$
are appropriately modified by partial saturation at higher
optical depths even though the relative intensities of the
line profile function are identical to the optically thin case.

From equations \ref{eq:RT} and \ref{eq:JbarJ} we can estimate the mean
radiation field, $\bar{J}$, for use in the statistical equilibrium
equations \ref{eq:SEHSE}. This completes the $\Lambda$ iteration.

In summary,
the {\it HSE} approximation 
is easily implemented in a standard ALI or Monte Carlo
code that models molecular rotational lines
simply by changing the line profile function.
We do not need to compute or store the hyperfine level populations.
We do not need to
model the radiative transfer of each hyperfine line
individually since the
hyperfine lines are included in the composite
line profile function of the rotational transition. 
Because the optical depth of the rotational lines are split among
their hyperfine components, the line trapping in the rotational
lines is approximately correct.

\section{The {\it Proportional} Approximation}

The {\it HSE} approximation is adequate if the hyperfine levels are
approximately in LTE even if the rotational levels are not.
However, observations sometimes find that the relative
intensities of the hyperfine lines 
do not correspond to those predicted by statistical equilibrium,
even for two lines with the same predicted intensities 
(Guilloteau \& Beaudry 1981; Caselli et al. 1995; 
Tafalla et al. 2002).  
In this case we can model the 
non-LTE excitation of the individual hyperfine lines by approximating
the collisional rate coefficients for the hyperfine transitions
rather than approximating the populations for the hyperfine levels.
The {\it proportional} approximation
assumes that the unknown rate for each collisional transition between
hyperfine levels is proportional to the known net rate between the
rotational levels and the statistical degeneracy of the
final hyperfine level (Guilloteau and Beaudry 1991; Daniel et al. 2006). 
The {\it proportional} approximation is 
computationally more demanding than the {\it HSE} approximation for two reasons.
First, the
number of levels in the statistical equilibrium equations now includes
the hyperfine levels. 
Second, the mean
radiation field and approximate Lambda operators must be determined
for each hyperfine line individually. The {\it proportional} approximation
generally results in greater accuracy, particularly for non-LTE
hyperfine emission.

The approximate collision rates for the hyperfine transitions are simply,
\begin{equation}\label{eq:hfapprox}
\tilde{C}_{JHJ'H'}
   = {g(J'H') \over g(J')}C_{JJ'}
\end{equation}
This definition
guarantees two requirements.
First, the average net collisional rate $C_{JJ'}$ between rotational levels
$J$ and $J'$ is equal to the weighted sum of the rates between the
hyperfine levels,
\begin{equation}\label{eqn:sumrate}
   C_{JJ'}= \sum_{HH'} {g(JH) \over g(J)}
   \tilde{C}_{JHJ'H'},  
\end{equation}
Second, the LTE populations, indicated by an asterisk, 
and collision rates between
any two levels
satisfy statistical equilibrium,
\begin{equation}
{{n^*_{JH}} \over {n^*_{J'H'}}}  = 
{{\tilde{C}_{J'H'JH}} \over {\tilde{C}_{JHJ'H'}}} 
= {{g_{JH}} \over {g_{J'H'}}} \exp ({h\nu / kT})
\end{equation}
where $h\nu$ = $\Delta E$ is the energy difference between the
levels.

With the approximate collision rates for all the transitions, we can
solve the statistical equilibrium equations for the populations of the
hyperfine levels,
\begin{eqnarray}\label{eq:SE}
&&\sum_{J>J'} n_{JH} A_{JJ'HH'}(1-\bar{\Lambda}_{JJ'HH'}) 
- (n_{J'H'}B_{J'JH'H} - n_{JH}B_{JJ'HH'})\bar{J}^{eff}_{JJ'HH'} 
\nonumber \\
&+&\sum_{J'>J} n_{J'H'} A_{J'JH'H}(1-\bar{\Lambda}_{JJ'HH'}) 
- (n_{JH}B_{JJ'HH'} - n_{J'H'}B_{J'JH'H})\bar{J}^{eff}_{JJ'HH'} 
\nonumber \\
&+&\sum_{J'H'}(n_{JH}C_{JJ'HH'} - n_{J'H'}C_{J'JH'H}) = 0\hfill
\end{eqnarray}
The effective mean radiation field is,
\begin{equation}\label{eq:Jbareff}
\bar{J}^{eff}_{JJ'HH'} = \bar{J}_{JJ'HH'} - \bar{\Lambda}_{JJ'HH'}S_{JJ'HH'}
\end{equation}
and the source function is,
\begin{equation}\label{eq:source}
S_{JJ'HH'} = {{j_{JJ'HH'}} \over {\alpha_{JJ'HH'}}}
\end{equation}
where the emissivity and opacity are,
\begin{equation}\label{eq:emissivityH}
j_{JJ'}(\nu) = \sum_{HH'} j_{JJ'HH'}\phi_{JJ'HH'}(\nu)
\end{equation}
\begin{equation}\label{eq:opacityH}
\alpha_{JJ'}(\nu) = \sum_{HH'} \alpha_{JJ'HH'}\phi_{JJ'HH'}(\nu)
\end{equation}
with $\phi_{JJ'HH'}(\nu)$ defined as in equation \ref{eq:phiH}.
These equations are essentially identical apart from notation to equations
\ref{eq:SEHSE}, \ref{eq:sourceHSE}, \ref{eq:emissivityJ}, and
\ref{eq:opacityJ}.
However, 
with this emissivity and opacity, the source function, even without 
the continuum, is no longer 
independent of frequency.

The radiation field and
the ALI operator
are computed slightly differently in the {\it proportional}
approximation than in the {\it HSE} case.
The mean radiation field is defined for each
individual hyperfine line so that equation \ref{eq:JbarJ} is replaced by
\begin{equation}\label{eq:JbarH}
\bar{J}_{JJ'HH'} = \int_\nu I_{JJ'}(\nu) \phi_{JJ'HH'}(\nu) d\nu
\end{equation}

The ALI operator 
is almost the
same as equation \ref{eq:aliJ}, but 
averaged separately over each individual 
hyperfine line profile (equation \ref{eq:phiH})
instead of over the summed profile (equation \ref{eq:sumprofile}).
\begin{equation}\label{eq:accelerationParameter}
\bar{\Lambda}_{JJ'HH'} = \int_\nu \bigg(1-\exp{\big(-\tau_{JJ'C}(\nu)\big)} \bigg)
r_{JJ'HH'C}
\phi_{JJ'HH'}d\nu
\end{equation}
and
\begin{equation}\label{eq:r291H}
r_{JJ'HH'C} = {{\alpha_{JJ'HH'}(\nu)}\over{\alpha_{JJ'}(\nu) + \alpha_C}}
\end{equation}
replaces equation \ref{eq:r291J}, with $\alpha_{JJ'}$ defined as in
equation \ref{eq:opacityJ}.
The radiative transfer solution is
defined the same way as in the {\it HSE} approximation
by equations \ref{eq:RT}, \ref{eq:sourceC}, and \ref{eq:tau}.


\section{Extrapolation to elastic rates}

Compilations of collisional rate coefficients for rotational transitions
generally do not include the elastic rates for transitions between the
same rotational level, $\Delta J = 0$, because
the forward and reverse rates are the same and therefore cancel out in the
equations of statistical equilibrium. However, hyperfine levels within
a rotational state can have different energies, and the forward and
reverse hyperfine collisional rates do not necessarily cancel even for 
transitions with $\Delta J = 0$. In order to estimate these hyperfine
collisional rates from equation \ref{eq:hfapprox}, we need to know the 
net rate for $\Delta J = 0$.

de Jong, Chu, \& Dalgarno (1975) suggested that collisional rates between
rotational levels could be parameterized by an equation of the form,
\begin{equation}
K_{JJ'} = a(\Delta J) {{g_J' } \over {g_J }} \bigg( 1 + {{\Delta E_{JJ'}}\over{kT}} \bigg)
\times \exp \bigg[ -b(\Delta J) \bigg( {{\Delta_{JJ'}} \over {kT}} \bigg)^{1/2} \bigg]
\end{equation}
where $a(\Delta J)$ and $b(\Delta J)$ are parameters to be determined.
This approximation is based on the assumption that all transitions with the
same $\Delta J$ are related because transitions which change the angular momentum
by $\Delta J$ are induced by the same term, $P_\lambda$, in the Legendre expansion
of the interaction potential,
\begin{equation}\label{eq:djcd75}
V(R,\Theta) = \sum_\lambda v_\lambda(R) P_\lambda (\cos \Theta)
\end{equation}
where $R$ and $\Theta$ are the separation and orientation of
the collision partners (Green \& Chapman 1978). 

\begin{figure}[t]
\includegraphics[width=5.25in,angle=90]{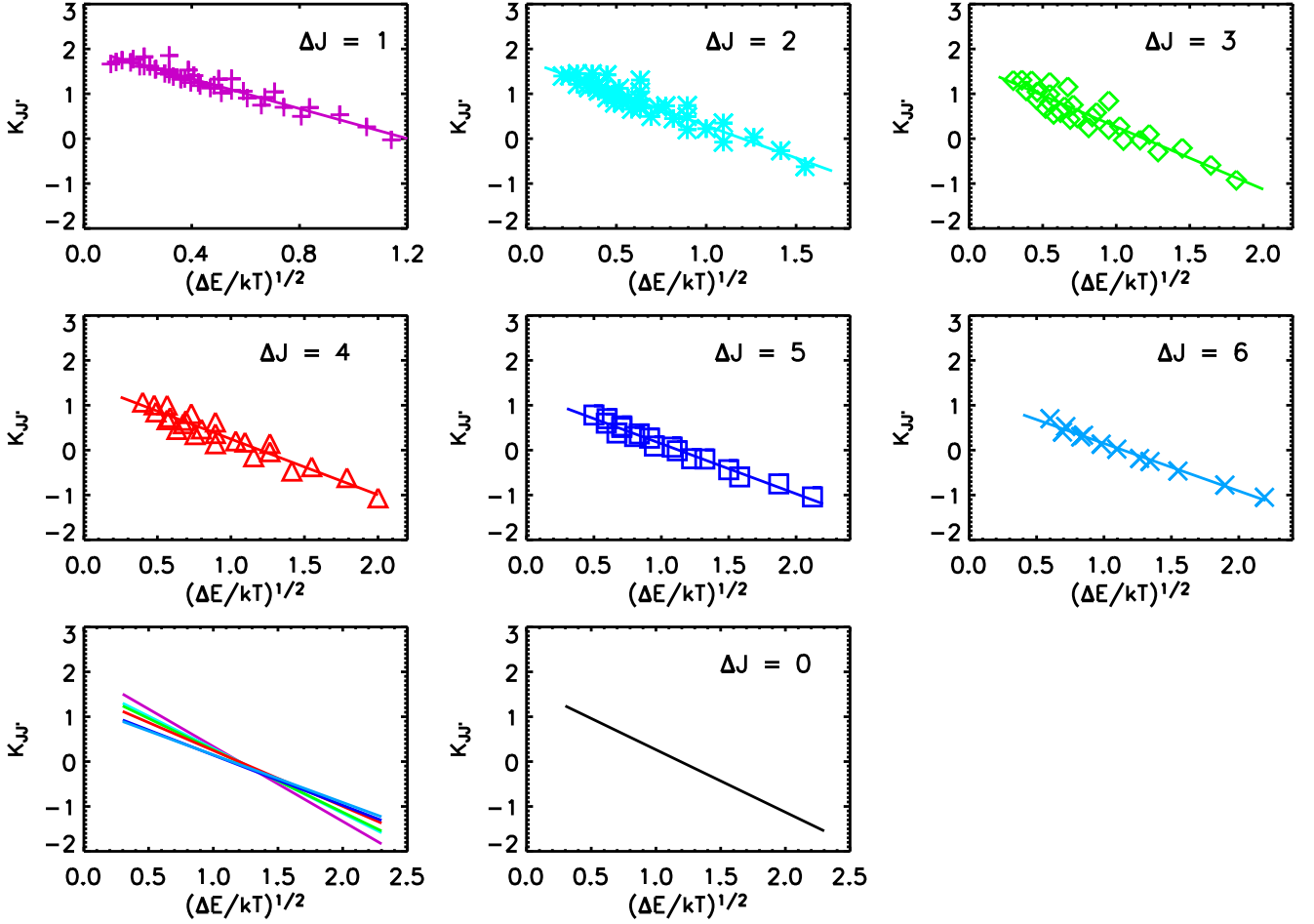}
\caption{Extrapolation to $\Delta J = 0$ by empirical fit
of the known inelastic collisional rates. The upper six
panels show the collisional rates $C_{JJ'}$ for six
different $\Delta J$ written
as $K_{JJ'} = \log \bigg(
{  {C_{JJ'} } \over { 1 + \Delta E_{JJ'} / kT  }} {{g_J   } \over {g_{J'}   }}
\times 10^{10}
\bigg) $ versus $(\Delta E / kT)^{1/2}$.
Equation \ref{eq:djcd75} is linear in this choice of coordinate axes
and is plotted for the six different $a(\Delta J)$ and $b(\Delta J)$ 
in each of the six panels. The lower two panels show these 6 lines
for $\Delta J = 1$ through 6 and the extrapolation to $\Delta J = 0$.
}
\label{fig:deltaJextrapolation}
\end{figure}

If we know a few rate coefficients, for example at a set of
temperatures, we can determine
the two parameters, $a(\Delta J)$ and $b(\Delta J)$
by a least-squares fit. Equation \ref{eq:djcd75} can then be used
to interpolate or extrapolate the rate coefficients as a
function of temperature. It turns out that the two 
parameters, $a(\Delta J)$ and $b(\Delta J)$, vary smoothly as
a function of $\Delta J$. Therefore, we can also use this equation to
extrapolate to transitions with different $\Delta J$, in particular
to $\Delta J = 0$. Figure \ref{fig:deltaJextrapolation} illustrates.
The symbols in the upper six panels show collision rates for transitions with
six different $\Delta J$. Here we use the collisional rates for 
HCO$^+$ (Flower 1999)
which should be similar to \diaz\ since both are molecular ions 
of about the same size.
The individual symbols represent the calculated rates for different temperatures.
From these known rates,
we find the parameters $a(\Delta J)$ and $b(\Delta J)$ for each $\Delta J$
by least-squares fits, one fit for each $\Delta J$. 
Lines representing equation \ref{eq:djcd75} for
each $\Delta J$ are shown in the six panels and shown together in the
lower right panel. From this collection of lines, or equivalently 
parameters $a(\Delta J)$ and $b(\Delta J)$ for $\Delta J = 1$ through $6$,
we can predict $a(\Delta J=0)$ and $b(\Delta J=0)$, shown in the last
panel. With this prediction for the elastic net rates we can use
equation \ref{eq:hfapprox} to predict the approximate hyperfine collision
rates for $\Delta J = 0$.

\section{Analysis of modeling}

\subsection{Comparison of {\it HSE} and {\it Proportional} approximations with observations}

\begin{figure}[t]
\includegraphics[width=5.25in,angle=90]{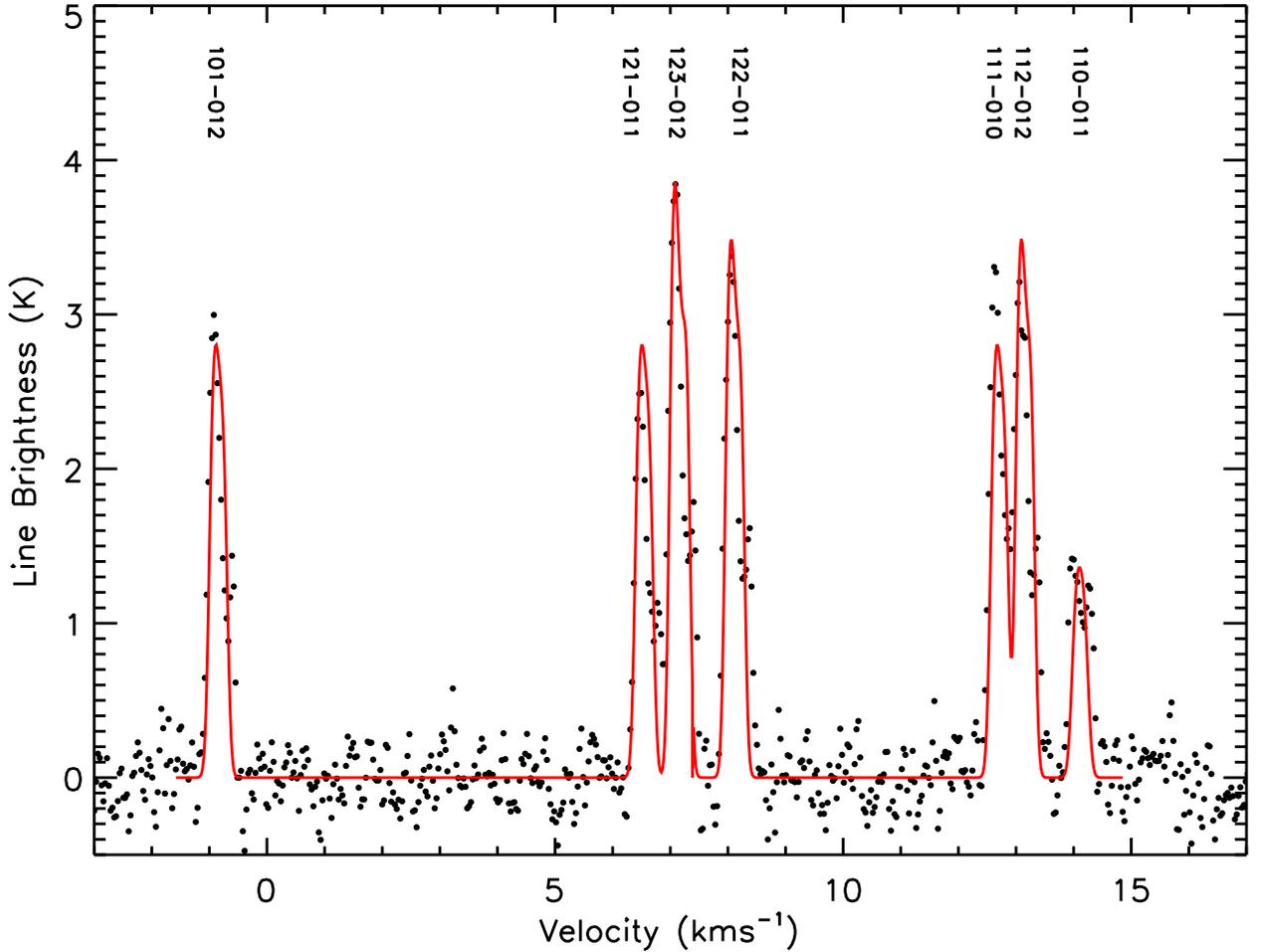}
\caption{Observed \diaz (1-0) spectrum of L1544 modeled with the {\it HSE} approximation. 
The dots show the observational data from Caselli et al.~(1995). The
line shows a model spectrum computed for a theoretical dark cloud
(Keto \& Caselli 2009) using our 3D radiative transfer code, MOLLIE,
and the {\it HSE} approximation. In this approximation, the three hyperfine
lines, 101--012, 121--011, and 111--010, necessarily have equal
relative intensities.
The velocity labeling includes the velocity of the L1544
cloud with respect to the Sun.
}
\label{fig:L1544LTE}
\end{figure}

\begin{figure}[t]
\includegraphics[width=5.25in,angle=90]{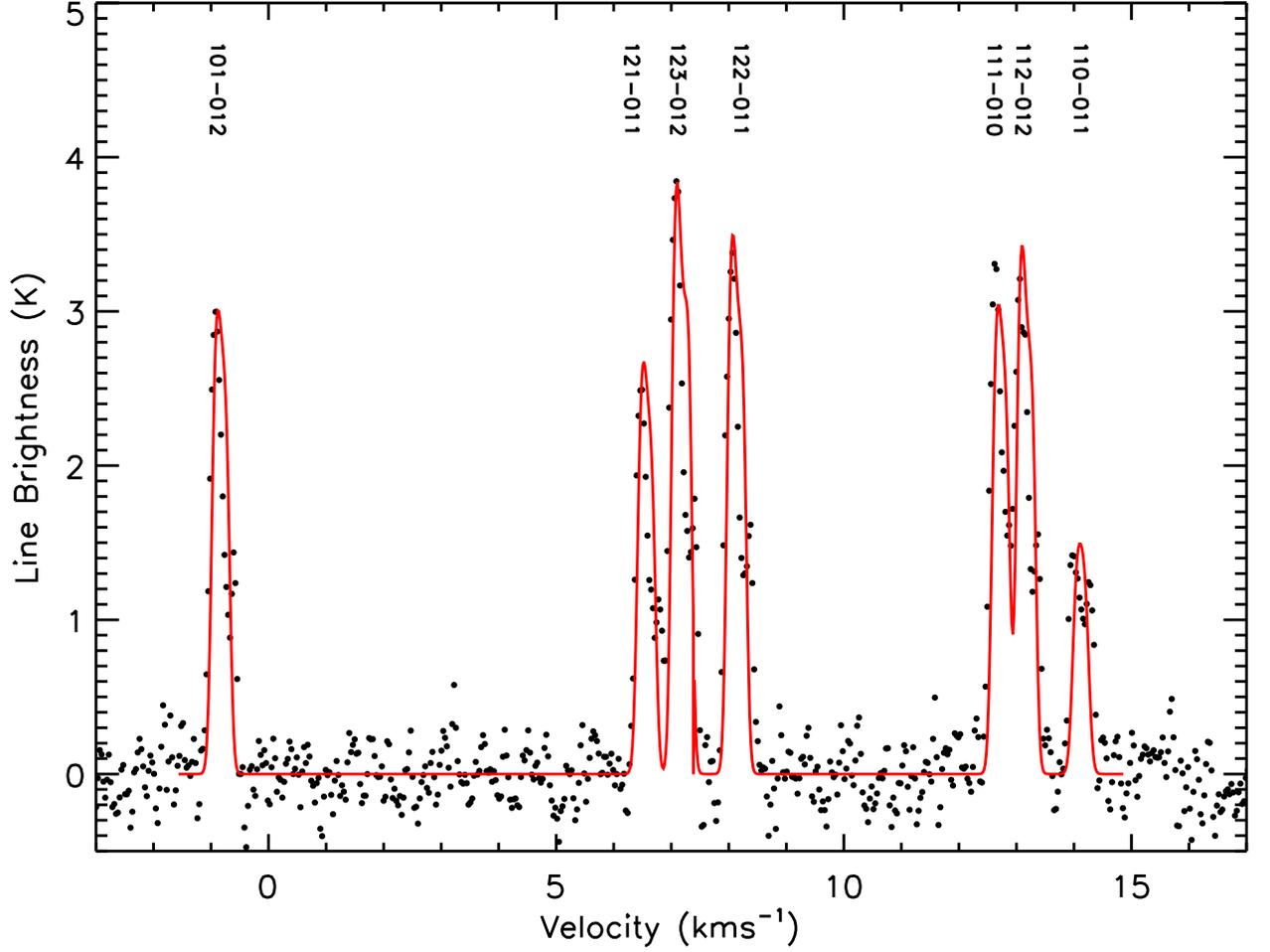}
\caption{Observed \diaz (1-0) spectrum of L1544 modeled with the {\it proportional} approximation. 
The same as figure \ref{fig:L1544LTE} except that the model spectrum
is computed with the {\it proportional} approximation, again using our
radiative transfer code, MOLLIE. Non-LTE excitation results in
unequal relative intensities for the 3 lines, 101--012, 121--011, and 111--010
and a better match to the data.
}
\label{fig:L1544nonLTE}
\end{figure}

Figures \ref{fig:L1544LTE} and \ref{fig:L1544nonLTE} compare
\diaz (1-0) spectra of the same model cloud computed 
using the {\it HSE} and {\it proportional} approximations against
the observed spectrum of L1544 (Caselli et al.~1999).
The model is taken from Keto \& Caselli (2009) and represents
a slowly contracting gas cloud in radiative equilibrium with
external starlight.  L1544 is thought to be an example of this
type of cloud.
These spectra were made with our 3D radiative transfer code,
MOLLIE (Keto 1990, Keto et al. 2004, Keto \& Caselli 2009). The {\it HSE}
approximation includes 8 rotational levels from $J=0$ to 7 and 
models the 7 $\Delta J = 1$ rotational lines. The hyperfine
splitting is included through the composite line profile
function (equation \ref{eq:sumprofile}). The {\it proportional}
approximation includes
64 hyperfine levels in the rotational levels $J=0$ through 7 and all
280 hyperfine lines between those hyperfine levels.  
The two approximations result in different
relative intensities for the hyperfine lines. The most evident are
the different intensities of the three lines 
$JFF_1 - J'F'F_1'$ = 101--012, 121--011, and 111--010.
In the LTE case, these three lines necessarily all have the same intensity
whereas with non-LTE excitation, the 121--011 hyperfine is
noticeably weaker and the 111--010 hyperfine is slightly
brighter. The {\it proportional} approximation represents a better match
to the data, yet for some purposes the {\it HSE} approximation may
be good enough. 

\begin{figure}[t]
\includegraphics[width=6.25in]{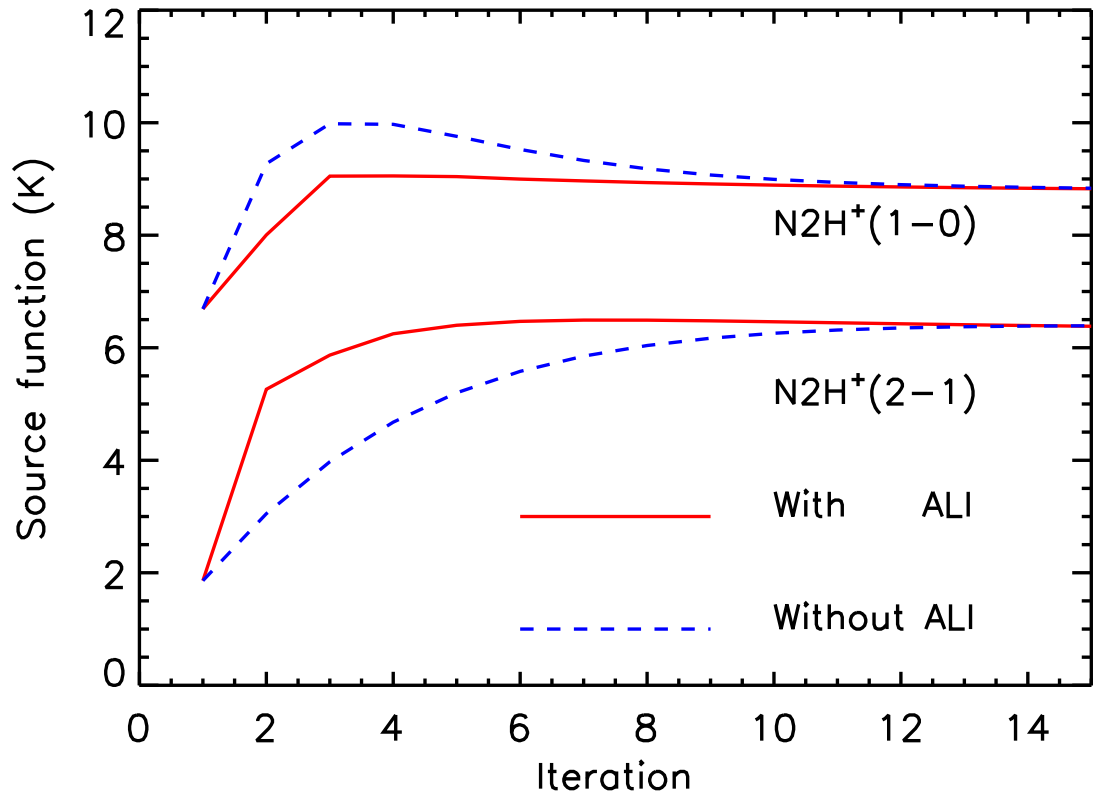}
\caption{The convergence of the $\Lambda$ iteration with and without acceleration.
The lines show the source function (equation \ref{eq:source}) 
of the main (123--012)
hyperfine line at the location of the center of the model cloud. The brightness of
the spectrum in figure \ref{fig:L1544nonLTE} is lower than the source
function because of
averaging lower brightness regions around the cloud center
within the observing beam. This figure shows that in this
calculation, the acceleration halves the required number of iterations.
}
\label{fig:convergence}
\end{figure}

Figure \ref{fig:convergence} compares the convergence of the
$\Lambda$ iteration in the {\it proportional} approximation
with the acceleration term
(equation \ref{eq:accelerationParameter})
and without ($\bar{\Lambda}_{JJ'HH'}=0$). In this example, the
optical depth is less than 10 and the $\Lambda$ iteration converges
quickly in both cases. However, convergence with the acceleration
requires half the number of iterations. At higher optical
depths, the acceleration would be considerably more significant.

\subsection{Comparison of ``exact" with ``approximate" collision rates.}

\begin{figure}[t]
\includegraphics[width=6.25in]{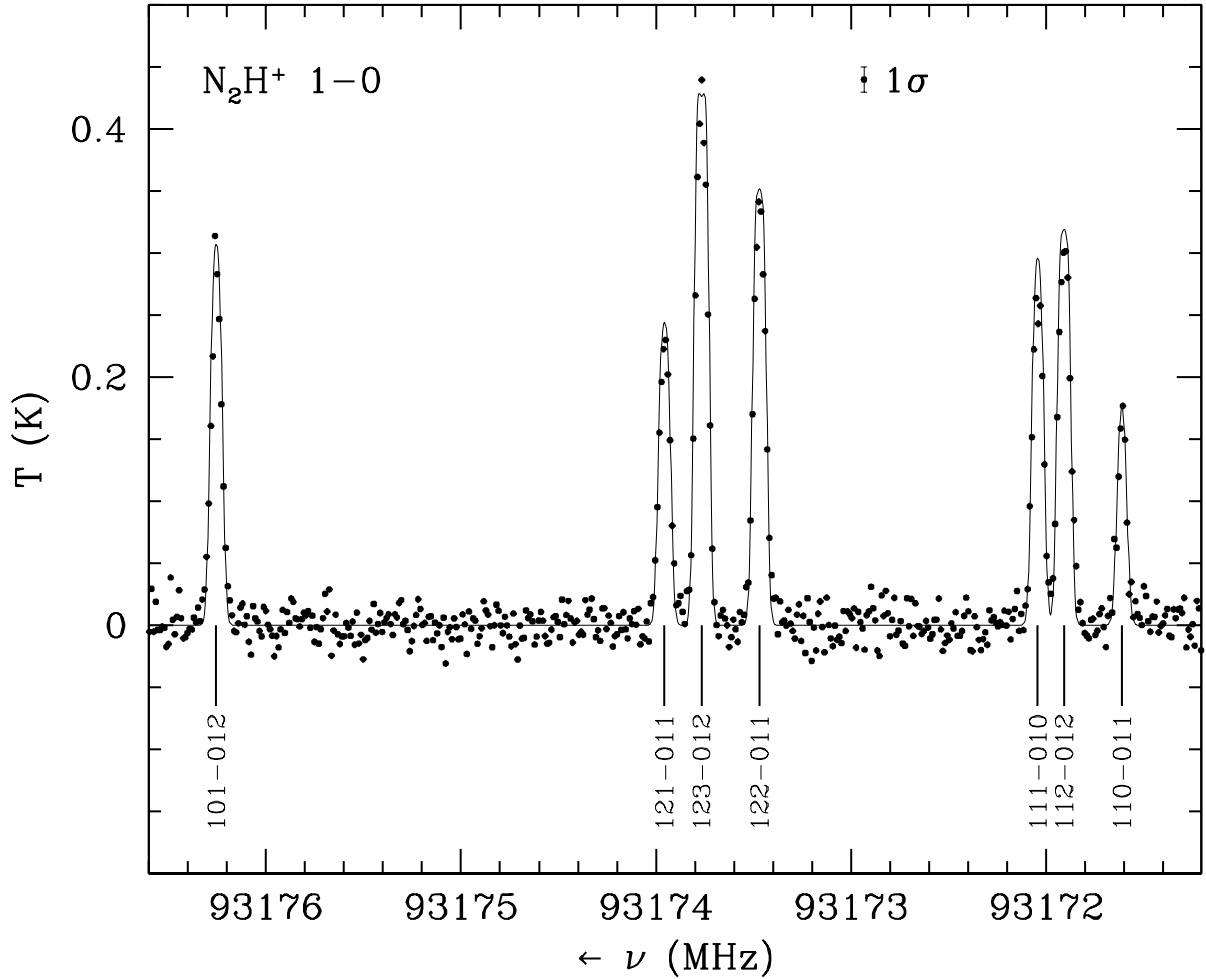}
\caption{Observed \diaz (1-0) spectrum of L1512 modeled using the ``exact" hyperfine 
collisional rates. The observational data (dots) are from 
Caselli et al.~(1995).  The line shows the model spectrum
computed using the ``exact" hyperfine collisional rates from 
Daniel et al.~(2006). The model spectrum is computed with a 1-dimensional
plane-parallel radiative transfer program.
}
\label{fig:L1512}
\end{figure}

Because the collisional rates for the hyperfine transitions of
\diaz\ have recently been calculated (Daniel et al. 2006),
we can compare the spectra computed with these rates 
and with the approximate 
collision rates of the {\it proportional} approximation.
In this comparison, we again use the same model for both calculations,
changing only the collisional rate coefficients.
In this example, 
we consider a uniform plane-parallel model of a molecular cloud with
a size of $4.11\times 10^{17}$ cm, density of $10^5$ cm$^{-3}$,
temperature of 8.9 K, abundance of \diaz\ relative to H$_2$ of
$3 \times 10^{-10}$, 
microturbulent line broadening of 0.06 kms$^{-1}$,
and a constant and zero velocity field.  The exterior boundary
condition assumes radiation at the 2.728 K background. 
These parameters were chosen to reproduce the \diaz (1-0) hyperfine line ratios
in the observations of L1512 (Caselli et al 1995). 
This calculation includes 
37 hyperfine levels in the rotational levels $J=0$ through 4 and all
145 hyperfine lines between those hyperfine levels.  
The fit to the data is shown in figure \ref{fig:L1512}. The
data for L1512 show the same pattern of non-LTE hyperfine line ratios
as for L1544 with the 121-011 hyperfine noticeably lower
than the 101--012 and 111--010 lines.

\begin{figure}[t]
\includegraphics[width=6.25in]{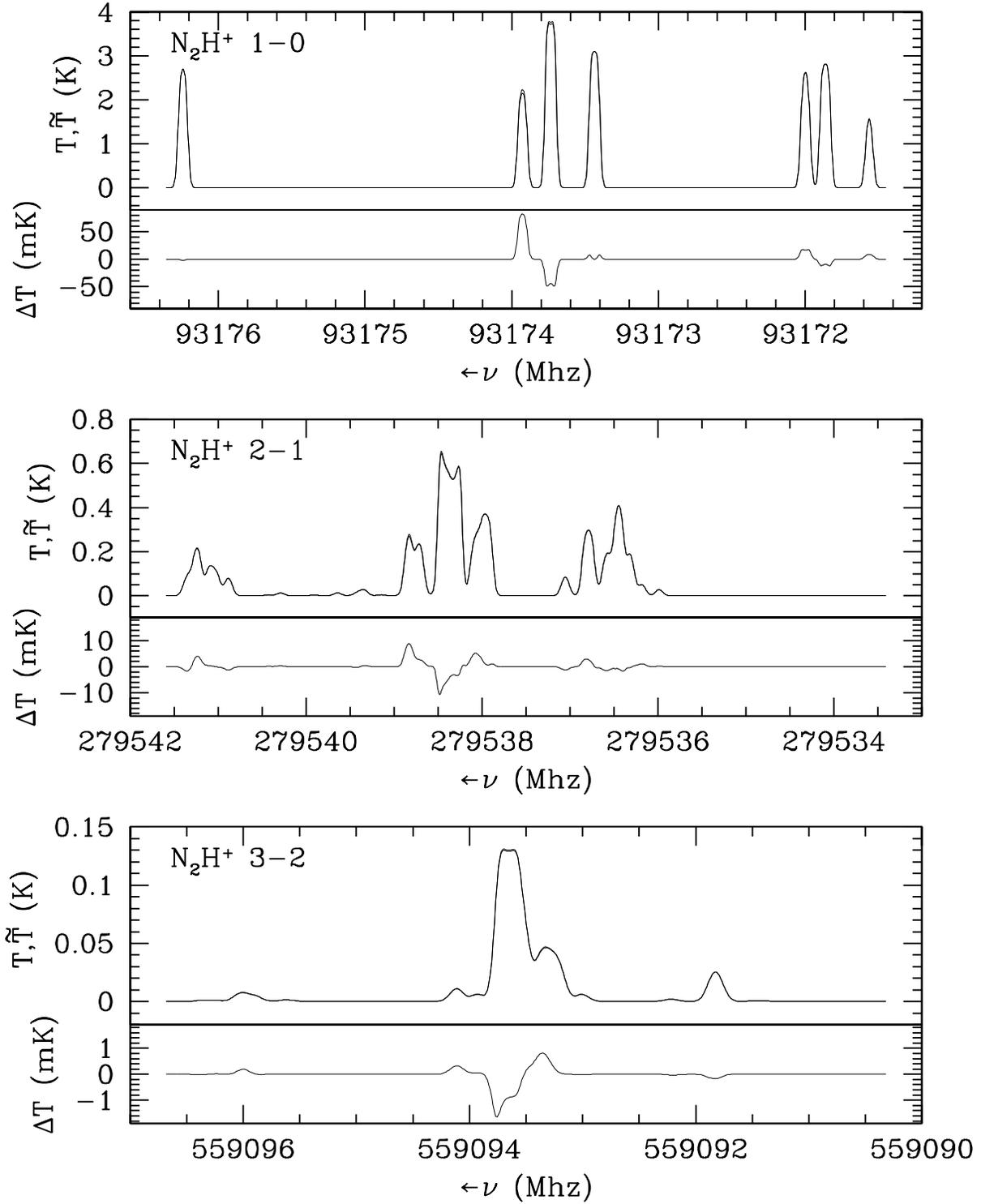}
\caption{Comparison of model \diaz\ spectra using approximate versus ``exact"
collisional rates. These 3 panels compare model spectra for the $\Delta J = 1-0$,
2--1, and 3--2 rotational transitions. The upper portion of each panel shows 
both the spectrum, $\tilde{T}$,
computed with the approximate hyperfine collisional rates of the {\it proportional} approximation
and the spectrum $T$ computed with the ``exact" rates from Daniel et al.~(2006). 
The two spectra are so close
as to be indistiguishable. The difference between the two spectra, $\Delta T$,
multiplied by 1000,
is plotted in the lower portion of each panel.}
\label{fig:rate_comparison}
\end{figure}

Figure \ref{fig:rate_comparison} 
compares the spectra computed from the approximate and ``exact"
collision rates. Spectra for the 3 lowest rotational
transitions of \diaz , (1-0), (2-1), and (3-2) are shown
along with the
difference between the two.  The
difference is less than one percent of the line strength. 
For most observations of radio frequency
molecular lines from dark clouds, this difference
would be below the typical signal-to-noise ratio. 
Based on this example, the {\it proportional} approximation is
adequate for \diaz\ and could be useful for other molecules
with unknown hyperfine collision rates.

\section{Conclusions}

The modeling of molecular spectra with hyperfine splitting by ALI or
Monte Carlo methods has been
hampered by the lack of collisional rate coefficients for the hyperfine
transitions. Two approximations previously suggested, the approximation
of hyperfine statistical equilibrium ({\it HSE}) 
and the {\it proportional} approximation, 
both provide satisfactory results in tests modeling \diaz\ spectra.
The {\it HSE} approximation, based on a modified line
profile function, is simpler to implement, faster to compute,
and models the non-LTE distribution in the rotational levels
but cannot model non-LTE distributions of the hyperfine levels 
themselves.
The {\it proportional} approximation uses easily computed 
approximate hyperfine collision rates, and is able to model
non-LTE hyperfine emission with an accuracy comparable to
calculations using the exact hyperfine collision rates.
These results suggest that these two methods could also be useful for
other molecules with hyperfine splitting.

\section{Appendix}

\subsection{Statistical Weights for \diaz } 

The hyperfine levels of \diaz\ are described by three angular momentum
quantum numbers, $J$, $F_1$, and $F$.  The first of these, $J$, refers
to the molecular rotation, which is coupled to the spins of the outer
and inner nitrogen nuclei $I_1=1$ and $I_2=1$, respectively.  The
coupling proceeds in two steps, first ${\hat F}_1={\hat J}+ {\hat
I}_1$, then ${\hat F}={\hat F}_1+ {\hat I}_2$, which provide the
remaining two quantum numbers $F_1$ and $F$.  The statistical weight
of a hyperfine level $JF_1F$ is given by $2F+1$, while the total
statistical weight of rotational level $J$ is
$g_{J}=(2I_1+1)(2I_2+1)(2J+1)= 9(2J+1)$.

In LTE, the population in hyperfine state $H = F_1 F$ relative to the
total population in rotational level $J$ is,
\begin{equation}\label{eqn:LTEpop}
n_{JH} = n_{J} \bigg( {{2F+1} \over {9(2J+1)}} \bigg )
\end{equation} 

The statistical degeneracies of the hyperfine states belonging
to each J level sum to the total statistical degeneracy of the J level.
\begin{equation}
\sum _F (2F+1) = 9(2J+1)
\end{equation}


\section{Einstein A for \diaz }

The Einstein A of a transition between rotational levels $JJ'$
is a weighted sum of all the Einstein A's between the individual
hyperfine states of each $J$ and $J'$ level, If the level
populations are in LTE, indicated by an asterisk,
\begin{equation}
n^*_J A_{JJ'} = \sum _{HH'} n^*_{JH} A_{JHJ'H'} R_{JHJ'H'}
\end{equation}
If the relative intensities $R_{JHJ'H'}$ are normalized 
so that,
\begin{equation} \label{eqn:normalization}
\sum _{HH} R_{JHJ'H'} = 1
\end{equation}
then
\begin{equation}
A _{JHJ'H'} = {{9(2J+1)} \over { {2F+1} A_{JJ'} R_{JHJ'H'}  }}
\end{equation}
For any rotational transition,
\begin{equation}
A_{JJ'} = { { 64 \pi ^4 \nu ^ 3} \over { 3 h c^3}} |\mu _{ij}|^2
\end{equation}

The average dipole moment,  $|\mu _{ij}|^2$, for a rotational
transition of a linear molecule is (Townes \& Schawlow, equation
1-76, pg 23),
\begin{equation}
 |\mu_{ij} |^2 = \mu^2 {{J} \over {2J+1}}
\end{equation}
if J is the initial state and the upper state. In this case,
$ J \rightarrow J -1$. As in Townes and Schalow, $|\mu_{ij} |^2$ can also be defined
in "absorption", $J \rightarrow J+1$, with $J$ as the
initial and lower state, or in "emission", $J+1 \rightarrow J$,
with J as the final and lower state. In these two alternate definitions,
$|\mu_{ij} |^2 = \mu^2 (J+1)/(2J+1)$ and 
$|\mu_{ij} |^2 = \mu^2 (J+1)/(2J+3)$ respectively.

With our definitions for $J$ and $R$, the Einstein A for a hyperfine transition is,
\begin{equation}\label{eqn:EinsteinA}
A_{JHJ'H'} = {{9(2J+1)} \over {2F+1}} \bigg [ {{ 64 \pi ^4 \nu ^ 3} \over { 3 h c^3}} \bigg ]
\mu ^2 {{J} \over {2J+1}} R_{JHJ'H'}
\end{equation}


The Einstein A's for the individual hyperfine transitions sum to,
\begin{equation}
\sum _{HH'} {{2F+1} \over {9(2J+1)}} A_{JHJ'H'} = A_{JJ'}
\end{equation}

\subsection{Collision rates for \diaz }

For \diaz\ the approximate collisional rate coefficients in the {\it proportional}
approximation are,
\begin{equation}\label{eqn:hfapproxDiaz}
\tilde{C}_{JF_1FJ'F_1'F'}
   = {2F'+1 \over g(J')}C_{JJ'}.   
\end{equation}

\begin{equation}\label{eqn:sumrateDiaz}
   C_{JJ'}= \sum_{F_1FF_1' F'} {2F+1 \over g(J)}
   \tilde{C}_{JF_1FJ'F_1'F'},  
\end{equation}

\subsection{Frequencies and relative intensities of \diaz hyperfine lines}

The frequencies and relative intensities of the hyperfine lines of \diaz\ 
are most accurately calculated by numerical methods (Pickett 1991). 
Dr. Luca Dore at the University of Bologna kindly supplied these
data. Table 1 shows the results for $JJ' = 1-0$ rotational transition. This
line is split into 16 hyperfine transitions at 7 different
frequencies to produce 7 hyperfine lines.
Tables 2 and 3 contain additional information on all the hyperfine states
and transitions for J levels 1 through 7.
Data on the frequencies and relative intensities of the hyperfine
transitions of N$_2$D$^+$ are available in \citet{Gerin2001} and
\citet{Dore2004}.

\clearpage

\begin{deluxetable}{ccccc}
\tablecaption{\diaz Hyperfine Line Data\tablenotemark{a}
     for $J=1 \rightarrow 0$}
\tablehead{
   \colhead{Line label\tablenotemark{b}} &
   \colhead{Frequency} &
   \colhead{Line Strength\tablenotemark{c}} &
   \colhead{Components} &
   \colhead{Component Strength\tablenotemark{c}}
}
\startdata
  $JF_1F$--$J'F'F_1'$ &  (MHz) & $S_{F_1F-F'F_1'}$ &
     $JF_1F \rightarrow J'F_1'F'$ & $S_{JF_1F \rightarrow J'F_1'F'}$\\[0.4em]
\tableline
\tableline
 110--011 & 93171.6086  & 0.33334&  $110 \rightarrow 011$ &0.33334 \\[0.4em]

 112--012 & 93171.9054  & 1.66667&  $112 \rightarrow 012$ &1.40832 \\
          &             &        &  $112 \rightarrow 011$ &0.25837 \\[0.4em]

 111--010 & 93172.0403  & 1.00000&  $111 \rightarrow 011$ &0.11979 \\
          &             &        &  $111 \rightarrow 012$ &0.37225 \\
          &             &        &  $111 \rightarrow 010$ &0.50797 \\[0.4em]

 122--011 & 93173.4675  & 1.66667&  $122 \rightarrow 011$ &1.40830 \\
          &             &        &  $122 \rightarrow 012$ &0.25836 \\[0.4em]

 123--012 & 93173.7643  & 2.33333&  $123 \rightarrow 012$ &2.33333 \\[0.4em]

 121--011 & 93173.9546  & 1.00000&  $121 \rightarrow 012$ &0.03938 \\
          &             &        &  $121 \rightarrow 011$ &0.64660 \\
          &             &        &  $121 \rightarrow 010$ &0.31402 \\[0.4em]

 101--012 & 93176.2527  & 1.00000&  $101 \rightarrow 010$ &0.17802 \\
          &             &        &  $101 \rightarrow 011$ &0.23361 \\
          &             &        &  $101 \rightarrow 012$ &0.58836 \\
\enddata
\tablenotetext{a}{Calculations by Luca Dore (private communication) using
the code of Pickett et al. (1991).}
\tablenotetext{b}{Each line is labeled by its strongest component.}
\tablenotetext{c}{Unit is $d^2$,
    where $d=4.3 \times 10^{-18}$ esu cm is the permanent
    dipole moment of \diaz.}

\end{deluxetable}

{}

\clearpage

\clearpage

\clearpage

\clearpage

\clearpage

\clearpage

\clearpage

\begin{deluxetable}{cccccc}
\tablecaption{\diaz Hyperfine Level Data}
\tablehead{
   \colhead{Level} & \colhead{Energy} & \colhead{Statistical Weight} &
   \colhead{J} & \colhead{F1} & \colhead{F}
}
\startdata
&       cm$^{-1}$ \\
\tableline
\tableline
    1       &0.0000    &5.0   &0   &1   &2 \\
    2       &0.0000    &3.0   &0   &1   &1 \\
    3       &0.0000    &1.0   &0   &1   &0 \\
    4       &3.1057    &1.0   &1   &1   &0 \\
    5       &3.1057    &5.0   &1   &1   &2 \\
    6       &3.1057    &3.0   &1   &1   &1 \\
    7       &3.1058    &5.0   &1   &2   &2 \\
    8       &3.1058    &7.0   &1   &2   &3 \\
    9       &3.1058    &3.0   &1   &2   &1 \\
   10       &3.1059    &3.0   &1   &0   &1 \\
   11       &9.3172    &5.0   &2   &2   &2 \\
   12       &9.3172    &7.0   &2   &2   &3 \\
   13       &9.3172    &3.0   &2   &2   &1 \\
   14       &9.3173    &7.0   &2   &3   &3 \\
   15       &9.3173    &9.0   &2   &3   &4 \\
   16       &9.3173    &5.0   &2   &3   &2 \\
   17       &9.3173    &3.0   &2   &1   &1 \\
   18       &9.3173    &5.0   &2   &1   &2 \\
   19       &9.3173    &1.0   &2   &1   &0 \\
   20      &18.6343    &7.0   &3   &3   &3 \\
   21      &18.6343    &9.0   &3   &3   &4 \\
   22      &18.6343    &5.0   &3   &3   &2 \\
   23      &18.6343    &9.0   &3   &4   &4 \\
   24      &18.6343   &11.0   &3   &4   &5 \\
   25      &18.6344    &7.0   &3   &4   &3 \\
   26      &18.6344    &5.0   &3   &2   &2 \\
   27      &18.6344    &7.0   &3   &2   &3 \\
   28      &18.6344    &3.0   &3   &2   &1 \\
   29      &31.0567    &9.0   &4   &4   &4 \\
   30      &31.0567   &11.0   &4   &4   &5 \\
   31      &31.0567    &7.0   &4   &4   &3 \\
   32      &31.0568   &11.0   &4   &5   &5 \\
   33      &31.0568    &7.0   &4   &3   &3 \\
   34      &31.0568   &13.0   &4   &5   &6 \\
   35      &31.0568    &9.0   &4   &5   &4 \\
   36      &31.0568    &9.0   &4   &3   &4 \\
   37      &31.0568    &5.0   &4   &3   &2 \\
   38      &46.5842   &11.0   &5   &5   &5 \\
   39      &46.5842   & 9.0   &5   &5   &4 \\
   40      &46.5842   &13.0   &5   &5   &6 \\
   41      &46.5842   &13.0   &5   &6   &6 \\
   42      &46.5843   & 9.0   &5   &4   &4 \\
   43      &46.5843   &11.0   &5   &6   &5 \\
   44      &46.5843   &15.0   &5   &6   &7 \\
   45      &46.5843   &11.0   &5   &4   &5 \\
   46      &46.5843    &7.0   &5   &4   &3 \\
   47      &65.2164   &13.0   &6   &6   &6 \\
   48      &65.2164   &11.0   &6   &6   &5 \\
   49      &65.2164   &15.0   &6   &6   &7 \\
   50      &65.2165   &15.0   &6   &7   &7 \\
   51      &65.2165   &11.0   &6   &5   &5 \\
   52      &65.2165   &13.0   &6   &7   &6 \\
   53      &65.2165   &17.0   &6   &7   &8 \\
   54      &65.2165    &9.0   &6   &5   &4 \\
   55      &65.2165   &13.0   &6   &5   &6 \\
   56      &86.9529   &15.0   &7   &7   &7 \\
   57      &86.9529   &13.0   &7   &7   &6 \\
   58      &86.9529   &17.0   &7   &7   &8 \\
   59      &86.9530   &17.0   &7   &8   &8 \\
   60      &86.9530   &13.0   &7   &6   &6 \\
   61      &86.9530   &15.0   &7   &8   &7 \\
   62      &86.9530   &19.0   &7   &8   &9 \\
   63      &86.9530   &11.0   &7   &6   &5 \\
   64      &86.9530   &15.0   &7   &6   &7 \\
\enddata
\end{deluxetable}
\clearpage

\begin{deluxetable}{cccccc}
\tablecaption{\diaz Hyperfine Level Data}
\tablehead{
   \colhead{Transition} & \colhead{Upper State} & \colhead{Lower State} &
   \colhead{Einstein A} & \colhead{Frequency} & \colhead{Relative Intensity}
}
\startdata
&&	&s$^{-1}$ &GHz \\
\tableline
\tableline
    1    &4   & 2  &3.6202E-05     & 93.1716086  &3.703754E-02  \\
    2    &5   & 2  &5.6121E-06     & 93.1719054  &2.870743E-02  \\
    3    &5   & 1  &3.0591E-05     & 93.1719054  &1.564799E-01  \\
    4    &6   & 3  &1.8390E-05     & 93.1720403  &5.644098E-02  \\
    5    &6   & 2  &4.3366E-06     & 93.1720403  &1.330977E-02  \\
    6    &6   & 1  &1.3476E-05     & 93.1720403  &4.136132E-02  \\
    7    &7   & 2  &3.0592E-05     & 93.1734675  &1.564777E-01  \\
    8    &7   & 1  &5.6123E-06     & 93.1734675  &2.870710E-02  \\
    9    &8   & 1  &3.6204E-05     & 93.1737643  &2.592588E-01  \\
   10    &9   & 2  &2.3410E-05     & 93.1739546  &7.184409E-02  \\
   11    &9   & 1  &1.4259E-06     & 93.1739546  &4.375910E-03  \\
   12    &9   & 3  &1.1369E-05     & 93.1739547  &3.489065E-02  \\
   13   &10   & 3  &6.4454E-06     & 93.1762527  &1.977944E-02  \\
   14   &10   & 2  &8.4583E-06     & 93.1762527  &2.595644E-02  \\
   15   &10   & 1  &2.1303E-05     & 93.1762527  &6.537287E-02  \\
   16   &11   &10  &2.7992E-08     &186.3401767  &4.474821E-06  \\
   17   &13   &10  &3.0605E-07     &186.3404984  &2.935571E-05  \\
   18   &16   &10  &1.5304E-06     &186.3424717  &2.446410E-04  \\
   19   &11   & 9  &1.1350E-05     &186.3424747  &1.814322E-03  \\
   20   &11   & 8  &2.6929E-05     &186.3426651  &4.304765E-03  \\
   21   &13   & 9  &2.0609E-04     &186.3427964  &1.976683E-02  \\
   22   &12   & 8  &1.7449E-04     &186.3429204  &3.905121E-02  \\
   23   &11   & 7  &6.8980E-05     &186.3429618  &1.102677E-02  \\
   24   &17   &10  &3.7407E-04     &186.3430541  &3.587865E-02  \\
   25   &12   & 7  &2.1961E-05     &186.3432171  &4.914732E-03  \\
   26   &18   &10  &3.8710E-04     &186.3432658  &6.187998E-02  \\
   27   &13   & 7  &3.3348E-05     &186.3432835  &3.198466E-03  \\
   28   &19   &10  &4.0897E-04     &186.3435179  &1.307500E-02  \\
   29   &11   & 6  &4.1510E-04     &186.3443890  &6.635387E-02  \\
   30   &14   & 8  &5.7200E-05     &186.3444527  &1.280094E-02  \\
   31   &11   & 5  &1.7271E-04     &186.3445240  &2.760721E-02  \\
   32   &13   & 6  &1.2012E-04     &186.3447107  &1.152066E-02  \\
   33   &14   & 7  &6.3721E-04     &186.3447494  &1.426027E-01  \\
   34   &16   & 9  &5.5131E-04     &186.3447698  &8.812664E-02  \\
   35   &12   & 5  &4.9863E-04     &186.3447793  &1.115900E-01  \\
   36   &13   & 5  &1.0558E-05     &186.3448457  &1.012611E-03  \\
   37   &15   & 8  &6.9509E-04     &186.3448501  &1.999999E-01  \\
   38   &16   & 8  &3.6017E-06     &186.3449601  &5.757276E-04  \\
   39   &13   & 4  &3.2467E-04     &186.3451424  &3.113882E-02  \\
   40   &16   & 7  &1.3303E-04     &186.3452569  &2.126449E-02  \\
   41   &17   & 9  &1.7064E-06     &186.3453521  &1.636638E-04  \\
   42   &18   & 9  &2.2819E-06     &186.3455638  &3.647582E-04  \\
   43   &18   & 8  &1.5809E-05     &186.3457542  &2.527066E-03  \\
   44   &19   & 9  &2.7376E-05     &186.3458160  &8.752108E-04  \\
   45   &17   & 7  &9.9051E-06     &186.3458392  &9.499830E-04  \\
   46   &18   & 7  &7.6819E-06     &186.3460509  &1.227927E-03  \\
   47   &14   & 5  &6.7821E-07     &186.3463116  &1.517738E-04  \\
   48   &16   & 6  &4.2661E-06     &186.3466841  &6.819109E-04  \\
   49   &16   & 5  &1.3617E-06     &186.3468190  &2.176538E-04  \\
   50   &17   & 6  &1.0864E-04     &186.3472664  &1.041905E-02  \\
   51   &17   & 5  &1.3927E-04     &186.3474014  &1.335727E-02  \\
   52   &18   & 6  &8.6729E-05     &186.3474781  &1.386316E-02  \\
   53   &18   & 5  &1.9549E-04     &186.3476131  &3.124782E-02  \\
   54   &17   & 4  &6.1497E-05     &186.3476981  &5.897943E-03  \\
   55   &19   & 6  &2.5875E-04     &186.3477303  &8.271997E-03  \\
   56   &20   &18  &3.4882E-07     &279.5085967  &1.028071E-05  \\
   57   &22   &18  &1.8603E-06     &279.5090304  &3.916298E-05  \\
   58   &22   &17  &3.2591E-06     &279.5092421  &6.861077E-05  \\
   59   &20   &16  &1.6539E-05     &279.5093908  &4.874558E-04  \\
   60   &20   &15  &3.3819E-05     &279.5095008  &9.967375E-04  \\
   61   &22   &16  &4.6438E-04     &279.5098245  &9.776190E-03  \\
   62   &21   &15  &4.3854E-04     &279.5098785  &1.661797E-02  \\
   63   &20   &14  &2.4117E-04     &279.5098982  &7.107855E-03  \\
   64   &21   &14  &3.6220E-05     &279.5102760  &1.372515E-03  \\
   65   &22   &14  &3.7711E-05     &279.5103319  &7.938893E-04  \\
   66   &25   &18  &2.4184E-06     &279.5110156  &7.127707E-05  \\
   67   &26   &18  &7.3142E-04     &279.5111328  &1.539760E-02  \\
   68   &26   &17  &2.2761E-03     &279.5113445  &4.791484E-02  \\
   69   &28   &19  &1.7593E-03     &279.5113847  &2.222227E-02  \\
   70   &23   &15  &1.8979E-04     &279.5114140  &7.191670E-03  \\
   71   &20   &12  &4.7599E-04     &279.5114305  &1.402838E-02  \\
   72   &27   &18  &3.2067E-03     &279.5115089  &9.451005E-02  \\
   73   &28   &18  &8.5742E-05     &279.5116369  &1.082997E-03  \\
   74   &20   &11  &3.0022E-03     &279.5116858  &8.848041E-02  \\
   75   &22   &13  &2.8961E-03     &279.5117978  &6.096633E-02  \\
   76   &21   &12  &3.2953E-03     &279.5118083  &1.248667E-01  \\
   77   &25   &16  &3.4109E-03     &279.5118097  &1.005275E-01  \\
   78   &23   &14  &3.5787E-03     &279.5118114  &1.356067E-01  \\
   79   &28   &17  &1.4792E-03     &279.5118486  &1.868338E-02  \\
   80   &24   &15  &3.7700E-03     &279.5118621  &1.746030E-01  \\
   81   &22   &12  &1.3014E-05     &279.5118642  &2.739705E-04  \\
   82   &25   &15  &5.2440E-06     &279.5119197  &1.545508E-04  \\
   83   &26   &16  &6.5715E-10     &279.5119269  &1.383404E-08  \\
   84   &22   &11  &3.5375E-04     &279.5121195  &7.446892E-03  \\
   85   &27   &16  &2.4855E-06     &279.5123029  &7.325300E-05  \\
   86   &25   &14  &3.4636E-04     &279.5123171  &1.020782E-02  \\
   87   &27   &15  &1.4795E-05     &279.5124130  &4.360447E-04  \\
   88   &28   &16  &1.9535E-05     &279.5124309  &2.467483E-04  \\
   89   &26   &14  &1.2137E-05     &279.5124343  &2.555042E-04  \\
   90   &27   &14  &7.1706E-06     &279.5128104  &2.113294E-04  \\
   91   &23   &12  &1.5528E-06     &279.5133437  &5.884003E-05  \\
   92   &25   &12  &2.0645E-06     &279.5138494  &6.084373E-05  \\
   93   &26   &13  &1.0267E-04     &279.5139002  &2.161342E-03  \\
   94   &26   &12  &1.0470E-04     &279.5139666  &2.203971E-03  \\
   95   &25   &11  &3.0217E-06     &279.5141047  &8.905449E-05  \\
   96   &26   &11  &5.4306E-04     &279.5142219  &1.143186E-02  \\
   97   &27   &12  &4.7716E-04     &279.5143427  &1.406264E-02  \\
   98   &28   &13  &2.8018E-04     &279.5144042  &3.538813E-03  \\
   99   &27   &11  &6.1682E-05     &279.5145980  &1.817842E-03  \\
  100   &28   &11  &1.4606E-04     &279.5147260  &1.844849E-03  \\
  101   &29   &27  &7.4736E-07     &372.6695840  &6.721026E-06  \\
  102   &31   &27  &2.3618E-06     &372.6700461  &1.651944E-05  \\
  103   &29   &25  &1.9884E-05     &372.6700772  &1.788133E-04  \\
  104   &29   &24  &3.7991E-05     &372.6701349  &3.416527E-04  \\
  105   &31   &26  &4.8732E-06     &372.6704221  &3.408610E-05  \\
  106   &31   &25  &8.3318E-04     &372.6705393  &5.827693E-03  \\
  107   &30   &24  &8.1806E-04     &372.6705729  &8.991692E-03  \\
  108   &29   &23  &5.2594E-04     &372.6705829  &4.729776E-03  \\
  109   &30   &23  &4.6530E-05     &372.6710209  &5.114248E-04  \\
  110   &31   &23  &4.0936E-05     &372.6710450  &2.863249E-04  \\
  111   &33   &27  &1.1668E-03     &372.6719768  &8.161053E-03  \\
  112   &35   &27  &2.4924E-06     &372.6720656  &2.241399E-05  \\
  113   &29   &21  &8.9202E-04     &372.6721184  &8.021803E-03  \\
  114   &32   &24  &4.1747E-04     &372.6721184  &4.588582E-03  \\
  115   &37   &28  &9.5314E-03     &372.6723529  &4.761915E-02  \\
  116   &33   &26  &9.9301E-03     &372.6723529  &6.945553E-02  \\
  117   &36   &27  &1.1416E-02     &372.6724208  &1.026622E-01  \\
  118   &33   &25  &1.5072E-07     &372.6724701  &1.054191E-06  \\
  119   &37   &27  &4.8486E-05     &372.6724809  &2.422388E-04  \\
  120   &29   &20  &1.0879E-02     &372.6724961  &9.783247E-02  \\
  121   &31   &22  &1.0756E-02     &372.6725246  &7.523414E-02  \\
  122   &30   &21  &1.1491E-02     &372.6725564  &1.262994E-01  \\
  123   &35   &25  &1.1675E-02     &372.6725589  &1.049958E-01  \\
  124   &32   &23  &1.1936E-02     &372.6725665  &1.311891E-01  \\
  125   &31   &21  &1.3873E-05     &372.6725805  &9.703108E-05  \\
  126   &34   &24  &1.2356E-02     &372.6725984  &1.604938E-01  \\
  127   &35   &24  &6.4340E-06     &372.6726165  &5.785998E-05  \\
  128   &37   &26  &1.9766E-03     &372.6728570  &9.875358E-03  \\
  129   &36   &25  &2.4223E-06     &372.6729140  &2.178349E-05  \\
  130   &31   &20  &7.0399E-04     &372.6729582  &4.924026E-03  \\
  131   &36   &24  &1.4411E-05     &372.6729717  &1.295961E-04  \\
  132   &37   &25  &1.7187E-05     &372.6729742  &8.586414E-05  \\
  133   &33   &23  &1.2905E-05     &372.6729758  &9.026414E-05  \\
  134   &35   &23  &6.6612E-04     &372.6730646  &5.990276E-03  \\
  135   &36   &23  &6.6651E-06     &372.6734197  &5.993804E-05  \\
  136   &32   &21  &2.2510E-06     &372.6741019  &2.474133E-05  \\
  137   &33   &22  &1.1135E-04     &372.6744553  &7.788442E-04  \\
  138   &33   &21  &8.9884E-05     &372.6745112  &6.286748E-04  \\
  139   &35   &21  &2.3130E-06     &372.6746001  &2.080038E-05  \\
  140   &33   &20  &1.0443E-03     &372.6748890  &7.304192E-03  \\
  141   &36   &21  &8.6344E-04     &372.6749552  &7.764664E-03  \\
  142   &37   &22  &6.7097E-04     &372.6749594  &3.352110E-03  \\
  143   &35   &20  &2.6493E-06     &372.6749778  &2.382391E-05  \\
  144   &36   &20  &5.2589E-05     &372.6753329  &4.729110E-04  \\
  145   &37   &20  &1.1084E-04     &372.6753931  &5.537304E-04  \\
  146   &38   &36  &1.0095E-06     &465.8220241  &3.636243E-06  \\
  147   &38   &35  &2.2212E-05     &465.8223792  &8.000664E-05  \\
  148   &38   &34  &4.0780E-05     &465.8223974  &1.468922E-04  \\
  149   &39   &36  &2.4047E-06     &465.8224922  &7.086887E-06  \\
  150   &39   &35  &1.3109E-03     &465.8228473  &3.863332E-03  \\
  151   &40   &34  &1.3104E-03     &465.8228727  &5.578354E-03  \\
  152   &38   &32  &9.2218E-04     &465.8228774  &3.321732E-03  \\
  153   &39   &33  &5.5265E-06     &465.8229362  &1.628706E-05  \\
  154   &39   &32  &4.3197E-05     &465.8233455  &1.273055E-04  \\
  155   &40   &32  &5.3971E-05     &465.8233527  &2.297488E-04  \\
  156   &42   &36  &1.7098E-03     &465.8243347  &5.038865E-03  \\
  157   &38   &30  &1.4184E-03     &465.8244229  &5.109088E-03  \\
  158   &41   &34  &7.4606E-04     &465.8244293  &3.175888E-03  \\
  159   &43   &36  &2.3110E-06     &465.8245447  &8.324153E-06  \\
  160   &42   &35  &2.8176E-07     &465.8246899  &8.303576E-07  \\
  161   &46   &37  &2.6930E-02     &465.8247706  &6.172843E-02  \\
  162   &42   &33  &2.7294E-02     &465.8247787  &8.043820E-02  \\
  163   &45   &36  &2.9420E-02     &465.8248172  &1.059706E-01  \\
  164   &46   &36  &3.5978E-05     &465.8248307  &8.246709E-05  \\
  165   &38   &29  &2.8443E-02     &465.8248609  &1.024497E-01  \\
  166   &39   &31  &2.8303E-02     &465.8248669  &8.341198E-02  \\
  167   &39   &30  &1.4222E-05     &465.8248910  &4.191243E-05  \\
  168   &40   &30  &2.9483E-02     &465.8248982  &1.255051E-01  \\
  169   &43   &35  &2.9740E-02     &465.8248999  &1.071222E-01  \\
  170   &41   &32  &3.0098E-02     &465.8249092  &1.281255E-01  \\
  171   &43   &34  &7.3139E-06     &465.8249180  &2.634444E-05  \\
  172   &44   &34  &3.0847E-02     &465.8249325  &1.515151E-01  \\
  173   &45   &35  &2.1966E-06     &465.8251724  &7.912265E-06  \\
  174   &46   &35  &1.6064E-05     &465.8251859  &3.682110E-05  \\
  175   &42   &32  &1.3213E-05     &465.8251880  &3.894043E-05  \\
  176   &45   &34  &1.4224E-05     &465.8251905  &5.123376E-05  \\
  177   &46   &33  &2.6025E-03     &465.8252747  &5.965332E-03  \\
  178   &39   &29  &1.1677E-03     &465.8253290  &3.441266E-03  \\
  179   &43   &32  &1.0930E-03     &465.8253980  &3.937066E-03  \\
  180   &45   &32  &6.1763E-06     &465.8256705  &2.224666E-05  \\
  181   &41   &30  &2.7699E-06     &465.8264548  &1.179111E-05  \\
  182   &42   &31  &1.1238E-04     &465.8267094  &3.311910E-04  \\
  183   &42   &30  &8.1702E-05     &465.8267335  &2.407777E-04  \\
  184   &43   &30  &2.3071E-06     &465.8269436  &8.309887E-06  \\
  185   &42   &29  &1.6357E-03     &465.8271716  &4.820332E-03  \\
  186   &46   &31  &1.1677E-03     &465.8272054  &2.676621E-03  \\
  187   &45   &30  &1.3564E-03     &465.8272160  &4.885510E-03  \\
  188   &43   &29  &2.4536E-06     &465.8273816  &8.837820E-06  \\
  189   &45   &29  &4.8153E-05     &465.8276541  &1.734440E-04  \\
  190   &46   &29  &9.4852E-05     &465.8276675  &2.174124E-04  \\
  191   &47   &45  &1.1935E-06     &558.9639040  &2.042074E-06  \\
  192   &47   &44  &4.2772E-05     &558.9641620  &7.318242E-05  \\
  193   &47   &43  &2.3919E-05     &558.9641765  &4.092575E-05  \\
  194   &48   &45  &2.2321E-06     &558.9643698  &3.231556E-06  \\
  195   &48   &43  &1.8969E-03     &558.9646423  &2.746241E-03  \\
  196   &49   &44  &1.9142E-03     &558.9646639  &3.779000E-03  \\
  197   &47   &41  &1.4288E-03     &558.9646653  &2.444574E-03  \\
  198   &48   &42  &5.8004E-06     &558.9648524  &8.397519E-06  \\
  199   &48   &41  &4.4831E-05     &558.9651311  &6.490390E-05  \\
  200   &49   &41  &5.9499E-05     &558.9651672  &1.174626E-04  \\
  201   &51   &45  &2.3611E-03     &558.9661589  &3.418204E-03  \\
  202   &47   &40  &2.0547E-03     &558.9662218  &3.515574E-03  \\
  203   &50   &44  &1.1782E-03     &558.9662309  &2.326019E-03  \\
  204   &51   &43  &2.9343E-07     &558.9664314  &4.248112E-07  \\
  205   &52   &45  &1.9404E-06     &558.9664518  &3.319926E-06  \\
  206   &54   &46  &5.9693E-02     &558.9666312  &7.070705E-02  \\
  207   &51   &42  &6.0051E-02     &558.9666414  &8.693816E-02  \\
  208   &54   &45  &2.9943E-05     &558.9666447  &3.546815E-05  \\
  209   &55   &45  &6.2917E-02     &558.9666677  &1.076489E-01  \\
  210   &48   &40  &1.4375E-05     &558.9666877  &2.081056E-05  \\
  211   &48   &39  &6.1233E-02     &558.9666949  &8.864964E-02  \\
  212   &47   &38  &6.1389E-02     &558.9666971  &1.050348E-01  \\
  213   &52   &44  &7.9800E-06     &558.9667098  &1.365354E-05  \\
  214   &49   &40  &6.2967E-02     &558.9667237  &1.243087E-01  \\
  215   &52   &43  &6.3299E-02     &558.9667243  &1.083022E-01  \\
  216   &50   &41  &6.3759E-02     &558.9667342  &1.258728E-01  \\
  217   &53   &44  &6.4941E-02     &558.9667527  &1.452991E-01  \\
  218   &54   &43  &1.5398E-05     &558.9669172  &1.823909E-05  \\
  219   &51   &41  &1.3346E-05     &558.9669202  &1.932111E-05  \\
  220   &55   &44  &1.4124E-05     &558.9669258  &2.416574E-05  \\
  221   &55   &43  &1.8145E-06     &558.9669402  &3.104611E-06  \\
  222   &54   &42  &3.3454E-03     &558.9671272  &3.962630E-03  \\
  223   &48   &38  &1.7433E-03     &558.9671630  &2.523815E-03  \\
  224   &52   &41  &1.6274E-03     &558.9672131  &2.784334E-03  \\
  225   &55   &41  &5.6603E-06     &558.9674290  &9.684427E-06  \\
  226   &50   &40  &3.1583E-06     &558.9682907  &6.234927E-06  \\
  227   &51   &40  &7.6533E-05     &558.9684767  &1.107989E-04  \\
  228   &51   &39  &1.1150E-04     &558.9684840  &1.614180E-04  \\
  229   &52   &40  &2.1317E-06     &558.9687696  &3.647241E-06  \\
  230   &51   &38  &2.3270E-03     &558.9689521  &3.368852E-03  \\
  231   &54   &39  &1.7712E-03     &558.9689698  &2.097963E-03  \\
  232   &55   &40  &1.9564E-03     &558.9689855  &3.347352E-03  \\
  233   &52   &38  &2.3211E-06     &558.9692449  &3.971297E-06  \\
  234   &54   &38  &8.5854E-05     &558.9694378  &1.016933E-04  \\
  235   &55   &38  &4.5549E-05     &558.9694608  &7.793205E-05  \\
  236   &56   &55  &1.3333E-06     &652.0931420  &1.217987E-06  \\
  237   &56   &53  &4.4263E-05     &652.0933151  &4.043571E-05  \\
  238   &56   &52  &2.5221E-05     &652.0933579  &2.304032E-05  \\
  239   &57   &55  &1.9271E-06     &652.0936022  &1.525759E-06  \\
  240   &57   &52  &2.5909E-03     &652.0938182  &2.051286E-03  \\
  241   &56   &50  &2.0449E-03     &652.0938368  &1.868063E-03  \\
  242   &58   &53  &2.6286E-03     &652.0938378  &2.721444E-03  \\
  243   &57   &51  &5.9171E-06     &652.0941110  &4.684698E-06  \\
  244   &57   &50  &4.6055E-05     &652.0942971  &3.646270E-05  \\
  245   &58   &50  &6.3735E-05     &652.0943595  &6.598698E-05  \\
  246   &60   &55  &3.1204E-03     &652.0953552  &2.470444E-03  \\
  247   &56   &49  &2.8009E-03     &652.0954039  &2.558698E-03  \\
  248   &59   &53  &1.7153E-03     &652.0954144  &1.775921E-03  \\
  249   &60   &52  &2.1814E-07     &652.0955711  &1.727079E-07  \\
  250   &61   &55  &1.3855E-06     &652.0957103  &1.265668E-06  \\
  251   &63   &55  &2.6453E-05     &652.0958303  &1.772095E-05  \\
  252   &63   &54  &1.1483E-01     &652.0958533  &7.692302E-02  \\
  253   &60   &51  &1.1519E-01     &652.0958640  &9.119825E-02  \\
  254   &57   &49  &1.4442E-05     &652.0958641  &1.143360E-05  \\
  255   &61   &53  &8.4931E-06     &652.0958833  &7.758603E-06  \\
  256   &64   &55  &1.1890E-01     &652.0958841  &1.086189E-01  \\
  257   &57   &48  &1.1654E-01     &652.0959001  &9.226698E-02  \\
  258   &56   &47  &1.1671E-01     &652.0959058  &1.066197E-01  \\
  259   &61   &52  &1.1935E-01     &652.0959262  &1.090254E-01  \\
  260   &58   &49  &1.1894E-01     &652.0959266  &1.231386E-01  \\
  261   &59   &50  &1.1991E-01     &652.0959362  &1.241465E-01  \\
  262   &62   &53  &1.2163E-01     &652.0959520  &1.407408E-01  \\
  263   &63   &52  &1.4947E-05     &652.0960462  &1.001292E-05  \\
  264   &60   &50  &1.3402E-05     &652.0960500  &1.061060E-05  \\
  265   &64   &53  &1.4074E-05     &652.0960571  &1.285665E-05  \\
  266   &64   &52  &1.2687E-06     &652.0961000  &1.158940E-06  \\
  267   &63   &51  &4.2008E-03     &652.0963391  &2.814175E-03  \\
  268   &57   &47  &2.4300E-03     &652.0963660  &1.923873E-03  \\
  269   &61   &50  &2.2692E-03     &652.0964051  &2.072937E-03  \\
  270   &64   &50  &5.0682E-06     &652.0965789  &4.629841E-06  \\
  271   &59   &49  &3.4558E-06     &652.0975032  &3.577825E-06  \\
  272   &60   &49  &7.2981E-05     &652.0976170  &5.777952E-05  \\
  273   &60   &48  &1.1019E-04     &652.0976530  &8.723968E-05  \\
  274   &61   &49  &1.8230E-06     &652.0979721  &1.665302E-06  \\
  275   &60   &47  &3.1221E-03     &652.0981189  &2.471810E-03  \\
  276   &63   &48  &2.4822E-03     &652.0981281  &1.662810E-03  \\
  277   &64   &49  &2.6639E-03     &652.0981459  &2.433460E-03  \\
  278   &61   &47  &2.2168E-06     &652.0984740  &2.025063E-06  \\
  279   &63   &47  &8.0111E-05     &652.0985940  &5.366651E-05  \\
  280   &64   &47  &4.3847E-05     &652.0986478  &4.005444E-05  \\
\enddata
\end{deluxetable}
\clearpage

\end{document}